\documentstyle[11pt,aaspp4]{article}
\begin{document}

\title{Disk-locking and the presence of slow rotators among solar-type stars 
in young star clusters}
\author{Sydney Barnes\footnote {Work partially completed at Lowell Observatory, Flagstaff, AZ and Yale University, New Haven, CT, USA }}
\affil{Department of Astronomy, Univ. of Wisconsin - Madison, WI, USA}
\author{Sabatino Sofia}
\affil{Department of Astronomy, Yale University, New Haven, CT, USA}
\author{Marc Pinsonneault}
\affil{Department of Astronomy, Ohio State University, Columbus, OH, USA}

\begin{abstract}

The simultaneous presence of both the so-called `ultra-fast rotators'
and slowly rotating stars among the solar-type stars 
($\sim 0.6-1.2 M_{\odot}$) in the same young star clusters has been a puzzle
in the field of stellar rotation. 
No model to date has been able to explain both by a single mechanism
intrinsic to the star and questions about the appropriate initial conditions 
for models often complicate the problem. 

In this paper, using the same starting conditions for the models that we used 
in examining the origin of the ultra-fast rotators in young star clusters, 
we show that the slowest rotators demand an extrinsic mechanism. 
Assuming that this mechanism is a disk-star interaction, we determine that 
a disk-locking timescale of a few Myr must operate for this type of star. 
If, instead of allowing (radial) differential rotation, we enforce solid-body
rotation, the models require timescales about two to three times as long.

\end{abstract}


\section{Background}

During the past decade, the study of rotation in solar-type stars has been one 
of the more active areas in the field of stellar evolution. 
Although our ultimate objective is to understand the consequences of rotation,
the first step requires that the models reproduce the observed rotation
periods/velocities of stars as a function of age.

It has been recognized for a while that the rotational properties of 
massive stars are quite different from those of solar-type stars (Slettebak
1956; Kraft 1967). The angular
momentum evolution of stars more massive than about 1.25 $M_{\odot}$,
approximately the point at which the surface convection zone which 
characterizes lower mass stars vanishes, is complicated. 
Endal and Sofia (1976, 1978, 1979) developed a method 
based on work by Kippenhahn and Thomas (1970), to model rotating stars in one 
dimension, using a series of nested, deformed shells and initially applied it 
to massive stars. 
The most recent work on massive star rotation that the authors are aware of 
is that by Heger, Langer \& Woosley (2000), again guided by the Endal and 
Sofia method. 
Suffice to say that, because of their multi-layered evolution, including 
convecting and semi-convecting shells, and their high rotation rates, 
modelling massive rotating stars is a formidable undertaking. 

Modeling solar-type stars, which we define for the purposes of this paper as 
those between about 0.6 and 1.2 $M_{\odot}$, has been comparatively easier 
because of the simpler structure of these stars. 
By including some scheme for angular
momentum loss through a magnetized wind, a number of groups have modeled 
rotating stars (eg. MacGregor \& Brenner 1991; Charbonneau \& MacGregor 1992;
Cameron, Campbell \& Quaintrell 1995; Bouvier, Forestini \& Allain 1997). 
These models emphasize either the interior or the wind. They
treat the internal structure in slightly different ways, 
making solid-body or two-zone models and treat the wind in various ways, 
some through a parameterization, 
and some by solving the wind equations. 
The Yale group (eg. Pinsonneault et al. 1989, Pinsonneault, Kawaler \& 
Demarque 1990; Chaboyer, Demarque \& Pinsonneault 1995; Barnes \& Sofia 1996; 
Sills, Pinsonneault \& Terndrup 2000) 
has continued solar-type star work with the Endal and Sofia scheme, 
using a large number of nested shells, with radial angular momentum
transfer and consequently radial differential rotation but no latitudinal
dependence of rotation, and using a Kawaler-type parameterization 
for the rate of angular momentum loss through the wind (Kawaler 1988).
Observations of rotation rates of these stars, available since the 1970s, 
have provided a solid observational underpinning to these studies. This is
the type of star that we will consider in this paper.

Nowadays, it is possible to study the rotation of even 0.1 and 0.2 $M_{\odot}$ 
stars in the Hyades and Pleiades open clusters (eg. Terndrup et al. 2000) 
and models have also been made for the very lowest mass stars, below about 
0.5 $M_{\odot}$ (Sills, Pinsonneault \& Terndrup 2000, hereafter SPT00),
which are fully convective, and evolve somewhat
differently from the solar-type stars. The hope is that the very simple
internal structure of these stars will allow a relatively isolated study of
the properties of the processes of angular momentum loss through their winds.

\section{Introduction}

Among solar-type stars, a significant problem centers 
on the necessity of accounting simultaneously for the presence of slow and 
rapid rotators in young star clusters given the small observed range in 
rotation 
rates among their pre-main sequence precursors.
This has proven to be a difficult proposition. The angular momentum evolution
of low mass stars is a function of the initial conditions, angular momentum
loss, internal angular momentum transport and now, disk-interaction; 
disentangling these effects has not been easy. For an extensive discussion of 
the observations and theoretical studies see Krishnamurthi et al. (1997), 
hereafter KPBS97, and for the extension to very low mass stars see SPT00.
In a previous paper (Barnes \& Sofia 1996, hereafter Paper I), we have
presented a scenario whereby one could account for the presence of the 
ultra-fast rotators (UFRs) observed in young star clusters. This involved
both starting the evolution at the stellar birthline, and changing the 
prescription for the rate of angular momentum loss.
Unlike the Kawaler (1988) formulation of angular momentum loss, we found it
necessary to assume that regardless of the initial stellar rotation 
rate\footnote{ We allowed periods ranging from 16d (the longest) to as short
as 2d at the birthline.}, the rate of angular momentum loss must saturate at 
high rotation speeds (e.g. MacGregor \& Brenner 1991). 
In Paper I, we also found that this saturation threshold must be mass 
dependent to explain the data; KPBS97 reached a similar conclusion in an 
extensive study that used a single 10d starting period.

Stauffer \& Hartmann (1987) first noted that the angular momenta of even the
slowest pre-main-sequence stars were much higher than those observed for slow 
rotators in young clusters. Since the angular momentum loss rate is expected, 
on theoretical grounds, to be smaller for slower rotators\footnote{ The loss 
rate varies as the rotation rate to a power between 1 \& 2; also see section 
5 of this paper.}, this posed a serious problem. 
Of the possible solutions suggested, perhaps the most promising 
concerns the effect that locking the young star to a surrounding disk would 
have on the subsequent rotational evolution of the star.
The idea of disk-locking is suggested by various observations of T Tauri star
rotation. Edwards et al. (1993) noted that the Classical T Tauri stars (CTTs) 
all have periods greater than 4 d with a most probable period of about 8.5 d, 
while the Naked T Tauri stars (NTTs) have a range of periods from 2$-$16 d, 
and 
include a significant number with periods shorter than 4 d. Because the CTT 
stars show evidence that they are surrounded by disks (cf. Attridge \& Herbst 
1992; Bouvier et al. 1993) and the NTT stars do not, a disk-locking scenario 
appears attractive. This paradigm suggests that stars start out as CTTs with 
periods of 4$-$16d, perhaps locked to their disks, and that these disks are 
eventually lost by accretion or otherwise, whereupon they become NTT 
stars. At the current time, it is not clear that this is true for all stars,
and Stassun et al. (1999) have questioned its validity for stars less massive
than 0.5 $M_{\odot}$. However, Herbst et al. (2000) point out that the T Tauri
period distributions are mass-dependent and reiterate that the one for 
$M > 0.25 M_{\odot}$ continues to display the previously noted bimodality.
Although the reality may eventually turn out to be 
considerably more complex, a simple consideration of the effects of and limits 
on disk-locking of young solar-type stars seems necessary.

Some theoretical work has been done along these lines. 
Konigl (1991), Cameron \& Campbell (1993) and Armitage \& Clarke (1996)
have demonstrated that star-disk interaction via a (usually dipole) 
magnetic field is capable of regulating the rotation rate of the central
star and indeed, of locking the star entirely, for field strengths in the
several hundred G to 1 kG range and plausible accretion rates and disk masses.
Bouvier (1994) and Bouvier \& Forestini (1994) considered the effect of 
disk-locking on solid-body models, and concluded that disk lifetimes of at 
least 10$-$20 Myr 
would be necessary to make the slow rotators. More recently,
Bouvier, Forestini \& Allain (1997) lowered the median
disk lifetime to 3 Myr but still required 10$-$20 Myr of disk-locking to 
explain the rotation rates of the slowest stars. 
This is somewhat large, in 
view of observational studies (Strom et al. 1989, 1990)
which suggest shorter disk lifetimes. 
Rotational velocity distributions have been calculated by 
Cameron, Campbell \& Quaintrell (1995) and Keppens, MacGregor \& Charbonneau 
(1995), that seem to reproduce many 
features of the observed distributions. However, the mass range of the 
models (only 1 $M_{\odot}$ models in the former and 0.8 $M_{\odot}$ and 
1 $M_{\odot}$ models in the latter) is limited. Also, comparisons with 
observations are made via distributions at fixed (open cluster) ages, without 
supplying tracks of the time evolution of rotation of the stellar models.

KPBS97 explored models with different assumptions about internal angular 
momentum transport: solid body (SB) models and differentially rotating (DR) 
ones with 
internal angular momentum transport by means of hydrodynamic mechanisms. They 
also considered different levels for the saturation of angular momentum loss 
at high rotation rates, and 
a range of disk lifetimes to produce the distribution of initial conditions.  
Modest disk lifetimes, of order 3 Myr, were needed to produce the slow 
rotators for the DR models, while in agreement with previous work, longer disk 
lifetimes (of order 20 Myr) were needed for the SB models.  However, for the
SB models, the slow rotators were found to exhibit little change in rotation 
during the early MS, while the DR models were found to spin down
significantly even at early ages.  The time dependence of the slow rotator 
phenomenon is therefore a good diagnostic of core/envelope coupling, and 
KPBS97 found that the data were more consistent with the DR models than the 
SB models. Because the pre-main sequence data indicate that there is an 
intrinsic range in the initial rotation rates in addition to any possible 
disk locking (KPBS97 used a single 10d starting period for all models), 
this could well influence the results. In addition, KPBS97 did not explicitly 
address the question of whether the data could be reproduced without disks.

More recently, SPT00 have extended this formalism to very low
mass stars, in view of the fact that rotation rates are becoming available
now for even the lowest mass stars in the Pleiades and Hyades (eg. Terndrup
et al. 2000). The same general formalism seems to hold for these very low 
mass, 
fully convective stars, in the sense that both magnetic saturation and disk 
locking seem to be necessary to account respectively for the fast and slow 
rotators down to the very lowest stellar masses. The prescription for angular
momentum loss can be simplified somewhat in this mass range, but some 
structural aspects which can be ignored in more massive stars must be 
considered. As in KPBS97, SPT00 chose to use a single starting period of 10d 
rather than a starting period range, in order to keep the parameter space 
tractable.

The main motivation of this work is to examine the behavior of models which
include both a range in initial rotation, and a range of disk locking 
timescales. The emphasis is on examining the conditions under which the
slowly rotating T Tauri stars can be made to evolve into the slow rotators 
observed in young star clusters, while simultaneously, the fast T Tauri stars 
evolve into the UFRs.
We first show that models without disks cannot reproduce the open cluster 
data, and then examine which combinations of disk lifetime and initial 
rotation rate can reproduce the rotation periods seen in open clusters.
Our goal is not to elucidate the nature of the star-disk interaction, but to 
understand the extent of disk-locking that would be necessary to account for 
the observed rotation periods of young cluster stars.
We examine the effects of both components of the initial conditions.
We do not evolve entire distributions in time because, in the opinion of the 
authors, the observations are still inadequately free of biases to test this
level of detail.
We wish to provide rotational evolution calculations for individual stars
in the 0.6$-$1.2 $M_{\odot}$ range (representing solar-type stars) that extend
and merge seamlessly with the models of Paper I that may be compared directly
to observations, especially those that are expected to become available soon
(cf. Barnes 2000; Mathieu 2000).

On the observational side, some confusion has been created by lumping 
together the observations of very low mass stars with those of
the solar-type ones. Also, the T Tauri periods are mostly for stars less 
massive than solar-type.
Because of the paucity of observations of very low-mass stars, this difference 
could not hitherto be addressed effectively. With the increase of the 
observational base, this distinction is not only possible but also necessary 
from now on. Although it is difficult to determine 
the masses of T Tauri stars, both Herbst et al. (2000) and Stassun et al.
(1999) have supplied estimated stellar masses with their T Tauri periods. 
This added information should improve
the situation considerably. Although complete consistency between the initial
conditions, masses, models and observations has not yet been attained, we hope
for such consistency soon. 

In Paper I (Barnes \& Sofia, 1996) we have demonstrated that the
UFRs (ultra-fast rotators) in young star clusters cannot be produced if the
angular momentum loss scheme that produces a Skumanich-type slowdown on the
main sequence were also valid on the pre-main sequence. We demonstrated that
they could be produced and indeed, fit the overall rotation period 
observations better, if one assumed that the magnetic field (and hence 
angular momentum loss) saturated beyond a threshold angular velocity. 
This has also been noted by MacGregor \& Brenner (1991), Cameron \& Li 
(1994) and Keppens et al. (1995). We now wish to consider 
whether the slower rotators in these clusters can be explained within the
same general framework. The following section (3) of this paper explains the 
framework that motivates this and our previous work. Section 4 discusses the 
observational constraints on the models. Section 5 describes the stellar 
models and the evolution code. Section 6 presents the results, followed by 
the concluding discussion in section 7.

  \section{The framework}
The framework we proposed in Paper I involved:
 \begin{enumerate}
  \item Beginning the evolution of solar-type stars at the deuterium main 
	sequence as suggested by Palla \& Stahler (1991). This is both 
	observationally
	justified since the birthline delineates the upper envelope of the
	T Tauri observations in the H-R diagram, and is theoretically
	appealing since it offers a uniform initial condition for stars
	of different masses. This is important because rotational 
	evolution is rapid on the Hayashi track. Starting off the models with 
	the observed rotation periods of T Tauri stars removes a number of 
	uncertainties that would otherwise contaminate the results and 
	complicate their interpretation.
  \item Incorporating some form of saturation of the stellar magnetic field
	or equivalently, of angular momentum loss beyond a threshold 
	rotation velocity which varies with stellar mass. There is some
	justification for this in the observed saturation of chromospheric
	emission (eg. Stauffer 1994) which is thought to scale with stellar
	magnetic fields. The point of saturation is approximately known to be
	of order $10 \Omega_{\odot}$, but has not been precisely constrained 
	as a function of stellar mass. Angular momentum 
	loss via a stellar wind is unable to account for the presence of the 
	UFRs in young star clusters regardless of the starting period unless 
	it incorporates saturation beyond a threshold angular velocity 
	(MacGregor \& Brenner 1991; Barnes \& Sofia 1996). 
	Furthermore, models
	that do not incorporate saturation cannot account for the observed
	dispersion in stellar rotation periods as a function of time, whereas
	those that include saturation can. 

Paper I also suggested that the scenario 
above would not be sufficient to account for the very slowest rotators 
observed in young star clusters and that some additional ingredient, perhaps 
disk-locking, might have to be invoked to explain them. Again, there is 
observational justification for this in the T Tauri data, as stated earlier. 
This has led to a picture in which young stars are born with disks to which
they seem to be locked in some way (since otherwise they should spin up as
they contract down the Hayashi track). At some subsequent stage in the 
evolution (which may vary from star to star), the disk is lost and the star 
is now free to evolve onward as an NTT star. 
This suggests the following addition to the framework:

  \item	Including some form of disk-regulated angular momentum loss for some
	period on the pre-main-sequence to account for the presence of the
	slow rotators.
 \end{enumerate}
This scenario is compelling in its simplicity. We now wish to test whether the
disk-locking scenario is indeed capable of explaining the slowest rotators
and if so, what sorts of disk lifetimes would be necessary. But before that,
we make a digression to discuss some observational constraints.

 \section{Observational Constraints}

Large samples  of spectroscopic $v \sin i$  data have been collected for 
solar-type stars in young open clusters for over a decade now, often in 
conjuction with measurements of lithium abundance in stars 
(eg. Soderblom et al. 1999 and references therein; Stauffer et al. 1997a and
1997b and references therein; Queloz et al. 1998). These helped define and 
clarify various issues related to the rotation of sun-like stars, but they were
particularly useful with respect to the UFRs because a star with a very large
$v \sin i$ value can only have a large angular velocity, regardless of the
angle of inclination, $i$. For slow rotators, $v \sin i$ data are less useful. 
For several years, only (fairly high) upper limits on $v \sin i$ were 
available for many of the slow rotators, and even if a small $v \sin i$ value
is measured, one cannot tell whether it arizes from a small $v$ or a small 
$\sin i$.
Moreover, uncertainties in stellar radii come into play when one wishes to
know the intrinsic angular rotation rate. Thus, for all rotators, and
especially for the slow ones, rotation periods, when available, are preferred.

\subsection{Rotation periods}
At present, there exists a modest but rapidly expanding database of rotation
periods of stars in young open clusters. Observations exist for T Tauri-type
stars (Attridge \& Herbst 1992; Edwards et al. 1993; Bouvier et al. 1993;
Bouvier et al. 1995; Choi \& Herbst 1996; Stassun et al. 1999; 
Herbst et al. 2000),
IC 2391 (Patten \& Simon 1996), 
IC 2602 (Barnes et al. 1999),
IC 4665 (Allain et al. 1996a),
Alpha Per (Stauffer et al. 1985; Prosser et al. 1993a, 1993b, 1995; O'Dell 
\& Cameron 1993; Allain et al. 1996b; Prosser \& Grankin 1997), 
the Pleiades (Van Leeuwen \& Alphenaar 1982; Van Leeuwen, Alphenaar \& Meys 
1987; Prosser et al. 1993a, 1993b, 1995; Krishnamurthi et al. 1998) and 
the Hyades (Radick et al. 1987; Prosser et al. 1995).
Although the data set is incomplete in various respects\footnote{The 
rotation period database, for example, is biased in favor of faster rotators
because of observing time constraints.} (see Barnes 2000 for a recent review), 
several conclusions relevant to this work may already be drawn from it:

 \begin{enumerate}
 \item	T Tauri stars in our mass range have periods ranging from $\sim $ 1 
	to 17 d. The NTT and CTT stars are believed to show a preference for 
	the shorter and longer period ends of the distribution, respectively.
 \item	The Hyades show a well-defined sequence of lengthening period with 
	increasing color until $B-V$ $\approx$ 1.2. However, in the region 
	beyond, there are some rapid rotators but as yet no evidence of the
	slow rotators that might be expected just by extending the solar-type
	observations to lower stellar masses\footnote{$V \sin i$ data from
	Terndrup et al. (2000) seem to confirm this paucity of slow rotators
	among M-type stars in the Hyades.}.
 \item	The young clusters IC 2391, IC 2602, Alpha Per and the Pleides have
	a number of UFRs (periods of 0.2$-$1 d). But they also contain a 
	number of slow rotators, including several in the 7$-$10 d range.
	This is remarkably long for clusters this young and creates a 
	rotational dispersion that is very difficult to explain simply.
 \end{enumerate}
It is this last group of observations of slowly rotating stars that concern 
us here. These slow rotators will be displayed henceforth in the plots of the 
stellar models (Figs.1$-$5) by the upper boxes whose x-width represents the 
age range of 30$-$120 Myr\footnote{The ages of young clusters are in a state of
flux, inconsistent across different dating techniques and hence across
clusters. The lithium ages (eg. Stauffer (2000) and references therein) are 
longer than the traditional ages (eg. Mermilliod (1981)). Our 30$-$120 Myr 
boxes are drawn wide enough to accommodate the range of these measurements, 
probably exaggerating the actual age spread. Moving them a little will not 
affect the gist of our conclusions.}.
The UFRs, which were the 
subject of Paper I, are represented by the lower boxes in the same figures.

\subsection{Initial periods}

It is probably worthwhile to ask whether this paradigm needs to be 
reconsidered in view of work by Stassun et al. (1999).
They have questioned the disk-locking paradigm because they
observe no correlation between infrared excess and rotation period. However,
we must remember that this dataset is almost entirely composed of stars 
below the mass range considered in this paper. Of the 254 rotation periods
reported in that paper, most have masses in the range of 0.15 to 0.4 
$M_{\odot}$. Of those with both masses and periods tabulated,
only 9 stars have masses in excess of 0.5 $M_{\odot}$. The period range 
of these 8 stars is 1.03d to 8.26d (which is the range shown in Figs. 1-5
using gray shading). Herbst et al. (2000) derive similarly short periods but
are also sensitive to, and indeed derive, much longer periods. 
Let us consider the two ends of this distribution.

 \begin{enumerate}
  \item The short period end: If we take the observations at face value and 1d
is indeed the shortest period for stars in this mass range at 1Myr ages,
this approximately works out to a 4d initial period for the masses under
consideration at the birthline, exactly the value we have chosen in this
and the previous paper to use as the starting point for the fast rotators.
  \item The long period end: The longest period in this mass range derived by
Stassun et al. (1999) is 8.26d, but this is due to the fact that they are not 
sensitive to longer periods. Indeed Herbst et al. (2000) find longer periods
(displayed as a cap on the Stassun et al. (1999) observations in Figs 1-5). 
Therefore, our choice of a 16d period to represent the slowest rotators is 
both reasonable and in agreement with the Herbst et al. (2000) data. 
 \end{enumerate}

It is remarkable that the range of the actual observations of Stassun et al. 
(1999) and Herbst et al. (2000) actually match the initial conditions we had 
chosen in 1996 (see figures) before masses were available for these stars. 
We conclude that our choices are still the appropriate ones to use for the 
stars in our mass range.

 \section{The stellar models and the evolution code}

All the stellar models used in this work begin at the deuterium main sequence,
at which point they are assigned zero age. These models were originally 
generated from polytropic ones higher up on the Hayashi track, which were
allowed to evolve downward in the H-R diagram (during which time their
internal structure stabilized), until they satisfied the mass-radius
relationship of Palla \& Stahler (1991). This step is important because, 
otherwise, the change in the moment of inertia of the star as the internal
structure stabilizes will influence the rotational evolution.

As in previous work, the code was calibrated (Helium abundance and mixing
length adjusted) such that a 1 $M_{\odot}$ model would reproduce the solar 
radius and luminosity at the solar age. We have chosen this age to be 4.54 
Gyr, adding 40 Myr to the canonical value to account for pre-main-sequence 
evolution. A choice of Z=0.01895 then leads to an H mass fraction of 0.708 
and a mixing length parameter of 1.758. These are identical to the parameters
used in Paper I.

We have evolved all the stellar models using the Yale Rotating Stellar 
Evolution 
Code (YREC). This code models the rotating stars using a series of nested,
deformed shells. It accounts for transport and redistribution of chemical
species, and of angular momentum within a star as a result of various
rotationally induced instabilities (cf. Pinsonneault et al. 1989, 1990). 
Convection zones are assumed to be
fully mixed and to rotate as solid bodies, so that instabilities are 
effective only in radiative regions of the star. Angular momentum loss
via a magnetic stellar wind is modeled by draining it from the outer 
convection zone via the parameterization:
  \begin{equation}
  \frac{dJ}{dt}=\cases
           {-K \Omega ^{1+4N/3} (\frac{R}{R_{\odot}})^{2-N}
              (\frac{\dot{M}}{10^{-14}})^{1-2N/3} 
              (\frac{M}{M_{\odot}})^{-N/3}, & when $\Omega < \Omega_t$;\cr
           -K(\Omega_t^{4N/3}) \Omega (\frac{R}{R_{\odot}})^{2-N}
              (\frac{\dot{M}}{10^{-14}})^{1-2N/3}
              (\frac{M}{M_{\odot}})^{-N/3}, & when $\Omega \ge \Omega_t$.\cr}
  \end{equation}
where $\Omega$, $R$ and $M$ represent the surface rotation rate, radius
and the mass of the star respectively, and the wind index, $N$, is set to 1.5.
This includes magnetic
saturation beyond a threshold rotation rate $\Omega_t$ (constant for a star
of particular mass). The rationale for
this expression is explained extensively in Paper I, but essentially, it is 
necessary to explain the presence of the ultra-fast rotators in young
star clusters. The actual values of the thresholds used are given later in
this section.

In addition, the present version allows us to lock the rotation rate of a 
young star on the Hayashi track for a specified time. This is done to mimic 
the effect of disk-locking on the pre-main-sequence. Solid body rotation is 
enforced during the time that disk-locking is in effect. This is not an 
unreasonable condition because YREC enforces solid-body rotation in the 
surface convection zone and because stars are almost entirely convective on 
the Hayashi track anyway.
Although this prescription does not model the interaction between the
star and the disk, its simplicity is attractive. We do not (yet) wish
to complicate the interpretation of the rotation period data with 
arguments about the exact shape and magnitude of the required magnetic 
fields, or the nature of the star-disk interaction. These issues have been
addressed, for instance, by Konigl (1991) and Cameron et al. (1995).
We merely wish to lock the rotation of the star for a while, and examine 
whether the same models that generated the fast rotators can also make the 
slow rotators with this one additional modification. This is one of the first 
steps and doubtless, improvements can and will be made.

Apart from the enforcement of disk-locking, the version of the
code used in this paper is identical to the one used in Paper I. 
In addition to the solar radius and luminosity calibration at 4.54 Gyr as 
mentioned above, the constant $K$ in equation (1) above was chosen such that
the 1 $M_{\odot}$ model with 10 day starting period and no disk-locking 
reproduces the solar rotation rate at the same age. 
The saturation thresholds,
$\Omega_{t}$ have been set to 2, 5, 8 and 12 $\Omega_{\odot}$ for the 0.6, 
0.8, 1.0 and 1.2 $M_{\odot}$ models respectively. These are the same values 
as were used in Paper I except that 
here we have also included a 1.2 $M_{\odot}$ model. 

For the solid-body models, it is necessary to modify the saturation thresholds.
If the thresholds are kept identical to the differentially rotating ones,
these models become unreasonably fast on the pre-main sequence and the 
inferred disk lifetimes would be even longer than the ones we quote here. We
have chosen instead to adjust the saturation thresholds for these models so
that the fastest ones account for the ultra-fast rotators, in symmetry with
the differentially rotating models. This threshold is then used for all the
solid-body models (of the same mass). The saturation thresholds used for the
solid-body models in this work are 5, 12.5, 20 and 30 $\Omega_{\odot}$ 
respectively for the 0.6, 0.8, 1.0 and 1.2 $M_{\odot}$ models. 
We remind readers that the difference
between the differentially rotating and solid-body models is that the former
include redistribution of internal angular momentum via hydrodynamical 
instabilities in radiative zones. Dynamical instabilities are treated by
instantaneous readjustment while secular instabilities are treated through
diffusion equations.

Further  details about the code may be obtained in 
Pinsonneault et al. (1989), Pinsonneault, Kawaler \& Demarque (1990), 
Chaboyer (1993), Chaboyer, Demarque \& Pinsonneault (1995), 
Krishnamurthi et al. (1997) and Pinsonneault \& Sofia (2000).
In summary, these models 
(1) begin at the birthline and include pre-main-sequence evolution, 
(2) use a parameterization of angular momentum loss that includes
saturation and 
(3) hold the rotation rate of the star constant for a specified period
on the pre-main-sequence, mimicking the effect of disk-locking.

 \section{Results}

As stated earlier,
we have generated a grid of stellar models of masses 0.6 $M_{\odot}$, 0.8
$M_{\odot}$, 1.0 $M_{\odot}$ and 1.2 $M_{\odot}$. The first set has been
evolved without any disk-locking and the others have been locked for  0.3, 1, 
3 and 10 Myr in turn. All the models begin their (pre-main sequence) evolution 
at the stellar birthline of Palla \& Stahler
(1991) with initial rotation periods of 4, 8.5 and 16 days.
The models have then all been evolved to Solar age (4.54 Gyr)
under the various conditions detailed hereunder. All of these models
incorporate a saturation of angular momentum loss with the
thresholds as indicated in the previous section.
We have also evolved an equivalent set of solid-body models of which only an 
illustrative selection are displayed here. However, all models are available 
in electronic form from the first author. 

 \subsection{Disk-free models}

The rotational evolution of (differentially rotating) models without 
disk-locking is displayed in Fig.1. Except for the 1.2 $M_{\odot}$ models 
which were not displayed there, these are the same models that we presented 
in Barnes \& Sofia (1996). 
The lower and upper boxes at 30$-$120 Myr represent observations of UFRs and
slow rotators respectively in young open clusters (IC 2391, IC 2602, Alpha Per
and the Pleiades). Given the current state of the observations, no attempt has 
been made to display the mass dependence, if any, of rotation in these 
observations. We also show, using gray bars at 1 Myr, the recent rotation 
period measurements in Orion in this mass range from Stassun et al. (1999), 
capped by the longer periods from Herbst et al. (2000). 

The figure shows that the range of periods we have used at the stellar 
birthline is generally adequate to understand the origin of the UFRs in
young open clusters in the context of models incorporating magnetic saturation
(as shown in Barnes \& Sofia 1996). Furthermore, these are also consistent 
with the fast rotators among the T Tauri stars in this mass range. The way
in which the faster rotators at earlier ages evolve into the UFRs among young
open clusters (and, as shown later, the faster rotators among older clusters)
is highly satisfactory, especially because the relevant period data are 
probably not biased against fast rotators.

The principal point we wish to make with this figure is that the stars 
rotating slowest at the birthline, with a starting period of 16 days, are 
unable to account for the slow rotators observed in the young open clusters. 
Neither are they capable of explaining the longer Orion rotation 
periods\footnote{To generate the slowest T Tauris this way would require
$\sim$80d periods on the birthline!}. 
This effect is primarily a consequence of the structural evolution of single 
stars on the pre-main sequence. 
The upshot of this is that the models are incapable of reproducing 
simultaneously the UFRs and the slow rotators. However, the T Tauri 
observations, as indicated earlier, suggest that many of the Orion stars
display circumstellar disk signatures and evidence that disk-interaction
acts to brake the central star. Thus, the following section considers the 
rotational consequences on these stars of varying amounts of disk-locking. 

There is another significant way in which these models are inadequate.
The rotational dispersion of the models beyond 1 Gyr is non-existent. 
Observations of nearby solar-type stars contradict this, as stars of similar
mass are found to have a range of rotation rates (eg. Baliunas, Sokoloff \& 
Soon 1996). This has its origin in the $\Omega$ dependence of the angular 
momentum loss rate and suggests that the dispersion observed may not entirely 
come from phenomena intrinsic to the star.

\subsection{Disk-locked models}

Fig.2 displays the rotational evolution of models that are disk-locked for 
0.3 Myr (dashed lines) and 1 Myr (solid lines). By `disk-locked,' we mean
that the rotation periods of the models are held constant for the indicated
times. The disk itself is not explicitly modeled in any way. Beyond the
indicated time, the 
models are allowed to evolve freely as detailed in section 5.
As expected, these models all rotate slower than the models without disks 
because, unlike the previous models, these have not been allowed to spin up 
during contraction on the pre-main sequence. 

The T Tauri data of Stassun et al. (1999) can be accounted for using only
0.3 Myr of disk-locking, while the longer periods of Herbst et al. (2000)
suggest that at least some of the slow rotators require the enforcement of 
1 Myr of disk-locking, i.e. locking from the birthline to Orion age. 
As regards the slow rotators in young clusters, many
of them (but perhaps not the slowest ones) can be accounted for using 
1 Myr of disk-locking. This suggests even longer disk-locking times for
some of these stars. In passing, we note that the models are able to 
retain some rotational dispersion beyond a Gyr.

Continuing in the same vein, Fig.3 displays the results of evolving models 
with disk-locking enforced for longer periods: 3 Myr (dashed lines) and 10 
Myr (solid lines). 
As is obvious from the figure, these models have no trouble accounting for
all the slow rotators. In fact, if one takes into account the mass-dependence
of rotation (not displayed in these figures), 3 Myr of locking may be 
sufficient.

Finally, Fig.4 displays the entire range of rotation periods that can 
presently be produced by these (differentially rotating) models. 
The lower and upper solid curves represent respectively the rotational 
evolution of disk-free stars with 4 d initial periods and of stars disk-locked 
for 10 Myr with 16 d initial periods. The dashed curve is for models with 
3 Myr of disk locking. They account reasonably well 
for both the UFRs and the slow rotators in young star clusters, with equivalent
models for the intermediate cases. 

In this figure we have also displayed the available rotation periods in the
250 Myr-old NGC 3532 (Barnes 1998) and the 600 Myr-old Hyades clusters 
(Radick et al. 1987). These show the limitations of these models in accounting
for the fast rotators in older clusters, especially for the more massive stars.
Biases against slow rotators in the observations do not presently allow us to
use them effectively to constrain disk-lifetimes, but it seems unlikely that
disk-lifetimes greater than 3 Myr would be necessary in any event.

These models also show that disk-interaction is an effective means of
retaining rotational dispersion among stars of similar mass beyond say, a Gyr,
as observed among old solar-type stars. Although we do not wish to test this
aspect of the models yet, given the limitations of both the models and the
observations, it is quite possible that the dispersion observed among old stars
owes its origin to disk-interaction in the first few Myr of its existence.
In summary, these models, although not completely consistent with all the
observations, reproduce the major features reasonably, and the periods of
young cluster stars in some detail.

\subsection{Solid-body models}

We have also evolved an equivalent grid of solid-body models. Some of these
are displayed in Fig.5 for illustrative purposes\footnote{All models are
available in electronic form from the first author.}. 
As mentioned earlier, we
have adjusted the saturation thresholds for the individual masses to ensure
that, as in the case of the differentially rotating models, the ones with
initial periods of 4 d reproduce the ultra-fast rotator observations. Thus,
by design, the faster rotating models are almost equivalent to the 
corresponding differentially rotating models. 

The difference between the two sets of models
(Figs.4 and 5) is striking for the disk-locked models. Whereas $\sim$3$-$10 
Myr of disk-locking sufficed to explain all of the slow rotators in the young 
clusters in the differentially rotating case, in the solid-body case 
disk-lifetimes of $\sim$10$-$20 Myr are necessary, suggesting disk-lifetimes 
2$-$3 
times longer than those for the corresponding differentially rotating models. 
In Fig.5, we have plotted the rotational evolution of disk-free solid-body 
models with 4 d initial periods (lower solid line) and of two SB models with 
initial periods of 16 d disk-locked for 10 and 20 Myr (dashed and solid lines 
respectively). This is the solid-body equivalent of Fig.4. 
Apart from requiring longer disk-lifetimes, these models are generally 
equivalent to the DR ones, and neither are very satisfactory in reproducing the
observations in older clusters.

There is another difference between the solid-body and the differentially 
rotating models. We draw attention to this difference in Fig.6 which
displays the behavior of the two kinds of models past 100 Myr.
The solid curves represent the later evolution of solid-body
models with 16 d initial periods that have been disk-locked for 20 Myr. If 
one agrees that this is a reasonable upper limit for disk-locking, then the
region above the solid line is inaccessible to such models and the discovery
of slow rotators in this region among older star clusters will be a problem
for them.
The differentially rotating models, on the other hand, do populate this zone,
as the dashed lines (representing differentially rotating models with an
initial periods of 16 d and disk-locking enforced for 3 and 10 Myr) show. 
We have also plotted the measured rotation periods in the Hyades (from
Radick et al. 1987 and Prosser et al. 1995) and NGC 3532 (from Barnes 1998). 
The Hyades observations
are consistent with both sets of models but the observations of NGC 3532
seem to be a bit troublesome for the solid-body models to accommodate. 
They also raise some questions about the lack of slow rotators in the Hyades,
an observational issue beyond the scope of this paper.

A note about the internal rotation rate of the Sun is perhaps in order. 
The inconsistency of the Yale rotational models with the internal rotational
profile of the Sun remains. The solar models rotate 5$-$20 times faster near 
the center at solar age than they do on the surface. This is at odds with the 
rotation profile of the Sun as determined by helioseismology (eg. Antia et 
al. 1998). To a certain extent, this is a problem. However, we are dealing
here with very young stars and it is likely that additional physics will have
to be invoked to flatten the profile (see MacGregor (2000) for a recent review
of possible solutions to this problem). It is also true that just
because the Sun is presently spinning close to solid-body does not necessarily
imply that it was {\it always} spinning as a solid-body. In fact, it seems
that stars must rotate differentially early-on in their evolution to match
the early rotational data and to be consistent with the disk-lifetime
calculations. The twin constraints of early differential rotation and later 
solid-body-type rotation suggest the existence of an angular momentum
transport mechanism that operates on timescales of hundreds of Myr (also see
SPT00). This
mechanism would bring the models into better agreement both with the Hyades
data and also might explain the flat solar rotation profile. Additional 
observations in older clusters would help clarify the situation considerably.

	\section{Conclusion and discussion}

We have demonstrated the relative impossibility of generating the slow 
rotators in young open clusters through a mechanism instrinsic to the star
within the framework that explains the origin of the ultra-fast rotators.
The slow rotators cannot be generated
from the slow-spinning T Tauri stars unless one takes recourse to
some extrinsic mechanism. Assuming that star-disk
interaction is responsible for the slow rotators, we find that disk-lifetimes
$\sim$ 3 Myr (for differentially rotating models) are required.

Thus, we need to do two separate things to explain the large observed 
dispersion in rotation rates of stars in young open clusters: both decrease
the angular momentum loss rate for fast rotators to make them even faster,
and couple the central star quite strongly for a few Myr to circumstellar
material in order to produce the slow rotators. This effectively splits
young solar-type stars into two groups, depriving us of a single, simple,
uniformly applicable way of treating young solar-type stars. If we allow
this, then why not also allow young stars to have different magnetic field
configurations? 
Although this is possible in principle, there is no corroborating data in
support of this idea. 
Thus, generating the rotational dispersion using disks is presently preferable
to resorting to differing magnetic field configurations. 

We find that for the differentially rotating models 
disk-locking for $\sim$ 3 Myr can account for almost all the presently 
available observations, with perhaps only a star or two requiring a longer 
locking period, but still less than 10 Myr. Solid body models require 
longer disk-locking times, 10$-$20 Myr. 
These latter timescales seem to be rather long. Even a 3 Myr timescale is
somewhat long, given that the T Tauri data suggest 1 Myr disk-lifetimes. 
However, it is notoriously difficult to place T Tauri stars in 
an H-R diagram in order to date them and the difference between 1 and 3 Myr 
might not be significant.

Although solid-body models seem to require unreasonably long disk-lifetimes,
it might not be appropriate to reject them entirely because helioseismic
inversions suggest a solar interior rotation rate closer to solid body than
to the fast interiors of the differentially rotating models (see previous 
section). A way to 
reconcile these problems might be just to postulate a radial angular momentum
transfer rate of say, a few hundred Myr, between the SB and DR cases. Indeed,
detailed observations in a sequence of open clusters might help to find this 
timescale empirically.

Lastly, we note that although the young open cluster rotation rates look like
they could be generated from the T Tauri rates, there might be difficulties
in explaining the rates in older clusters like the Hyades and even in the
younger cluster NGC 3532. In particular, our models are probably not treating
the mass-dependence of the rotation rate correctly since, for example, the 
$1.2 M_{\odot}$ models are rotating slower than the observations in the Hyades.
The observations themselves are not consistent. There are slower rotators in
the younger NGC 3532 cluster than there are in the Hyades, raising the issue
of whether the Hyades data are biased and how much so. Perhaps a more complete
period census of the Hyades is needed and maybe one of Praesepe as well.
These issues, both observational and theoretical, will we hope, be addressed 
and understood in the near future.

{\it Acknowledgements.} 
SB would like to acknowledge the B.F. Foundation for funding his tenure at
Lowell Observatory through the Lowell Fellowship, the McKinney Foundation
and the NSF for support under AST-9731302 at the Univ. of Wisconsin. Many of 
these results were obtained as part of SB's PhD dissertation work at Yale 
University, which supported him for several years through a student fellowship.

\clearpage
 
\begin{figure}[1]
\plotfiddle{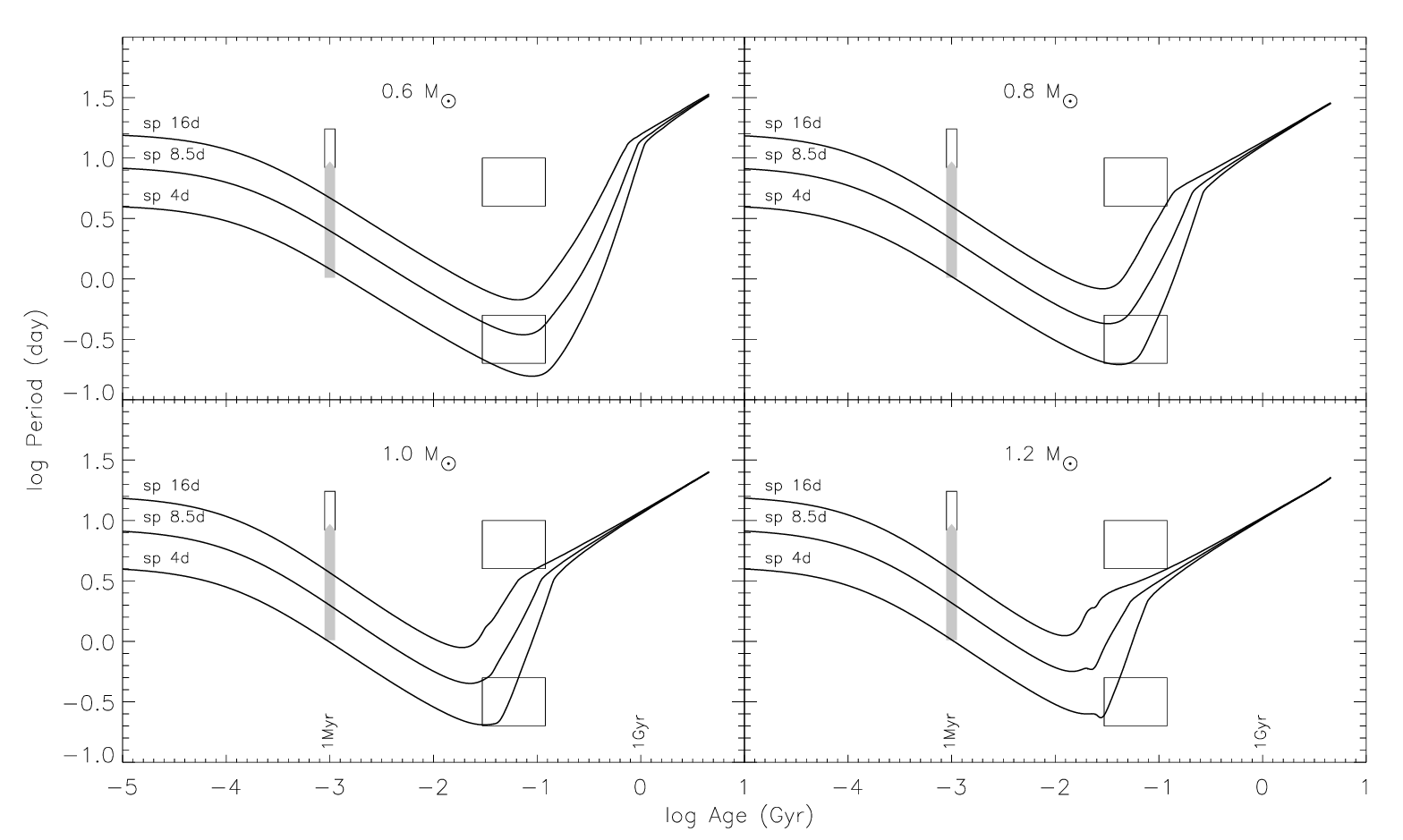}{6.5in}{0}{85}{85}{-290}{-150}
\caption[fig1.ps]{Rotational evolution of stellar models without disk-locking. The lower and upper boxes represent the UFRs and the slow rotators respectively in Alpha Per and the Pleiades. The letters `sp' indicate the starting period of the models. The gray bar at 1 Myr represents the rotation period measurements available in this mass range from work by Stassun et al. (1999) capped by longer periods from Herbst et al. (2000). Note that their fastest rotators are matched well by these disk-free models.}\label{fig1}
\end{figure}
 
\begin{figure}[2]
\plotfiddle{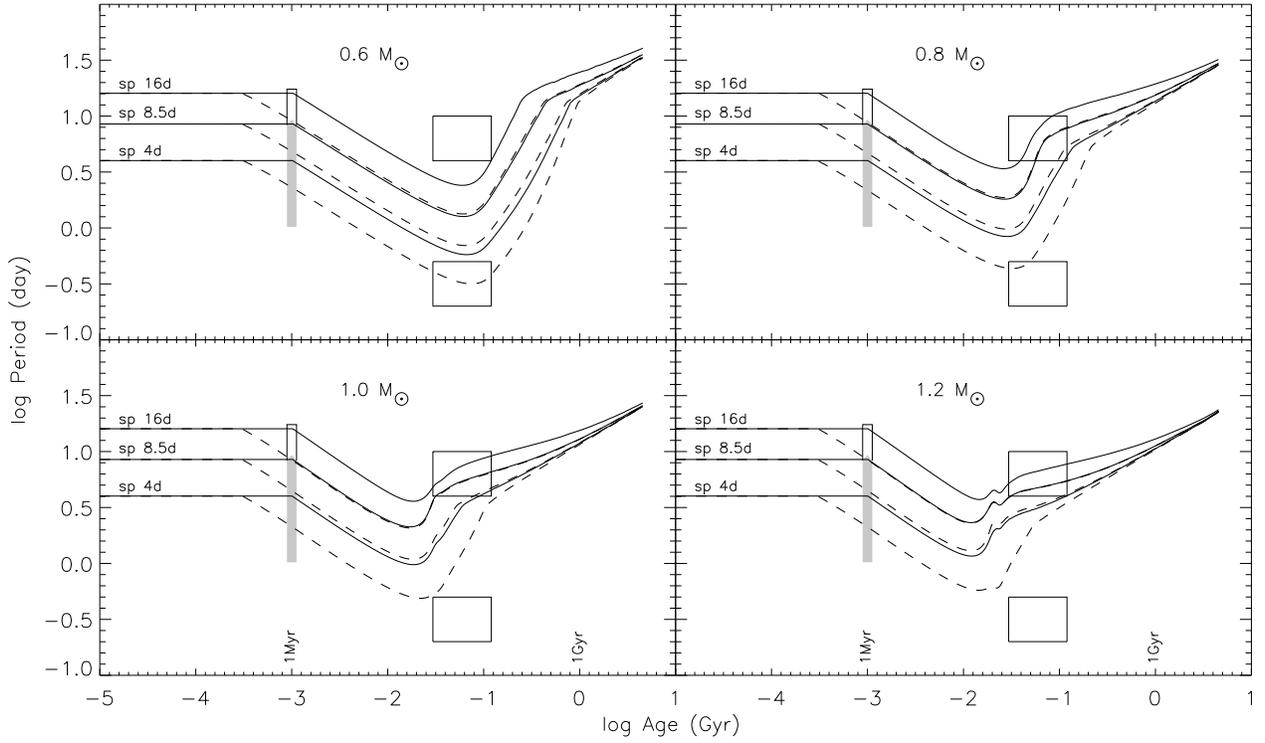}{6.5in}{0}{85}{85}{-290}{-150}
\caption[fig2.ps]{Rotational evolution of stellar models with disk-locking enforced for 0.3 Myr (dashed lines) and 1 Myr (solid lines). The observations plotted are the same as in Fig.1. Note that modest disk-locking can account for many but perhaps not all of the slow rotators in young star clusters.}\label{fig2}
\end{figure}
 
\begin{figure}[3]
\plotfiddle{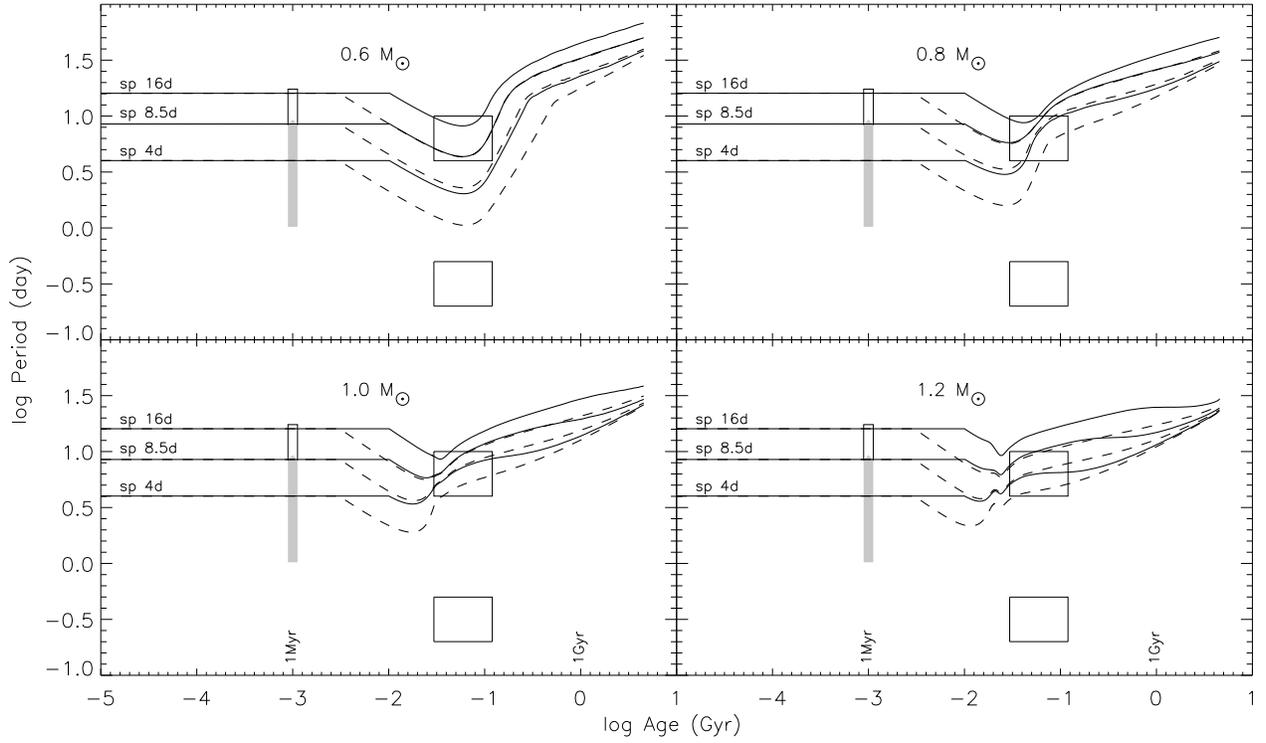}{6.5in}{0}{85}{85}{-290}{-150}
\caption[fig3.ps]{Rotational evolution of stellar models with disk-locking enforced for 3 Myr (dashed lines) and 10 Myr (solid lines). The observations plotted are the same as in figures 1 and 2. Note that these models can account for all the slow rotators observed in young star clusters.} \label{fig3}
\end{figure}
 
\begin{figure}[4]
\plotfiddle{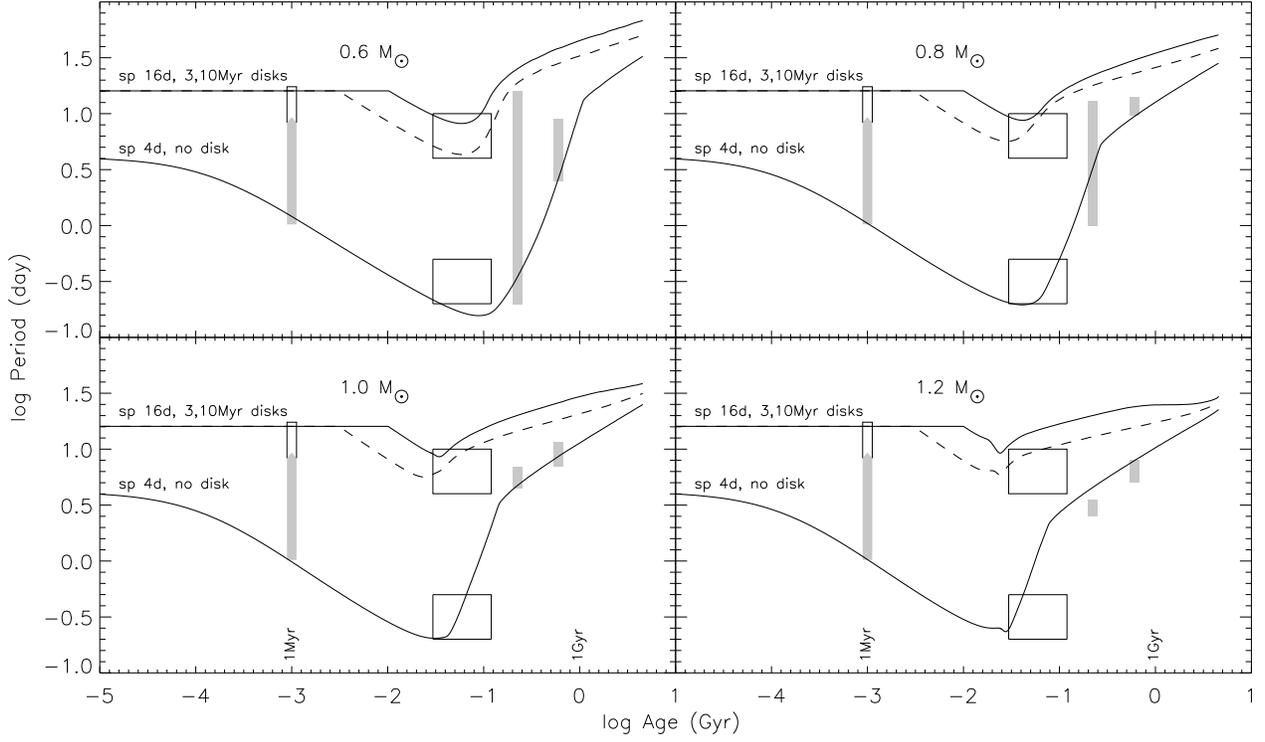}{6.5in}{0}{85}{85}{-290}{-150}
\caption[fig4.ps]{The range in stellar rotation period produced by differentially rotating models without disks and those disk-locked for 3 and 10 Myr. Note that these models can explain essentially the entire range of rotation period observations among solar-type stars in young star clusters. In addition to the T Tauri and young open cluster data, we have added observations in NGC 3532 (Barnes 1998) and the Hyades (Radick et al. 1987).}\label{fig4}
\end{figure}

\begin{figure}[5]
\plotfiddle{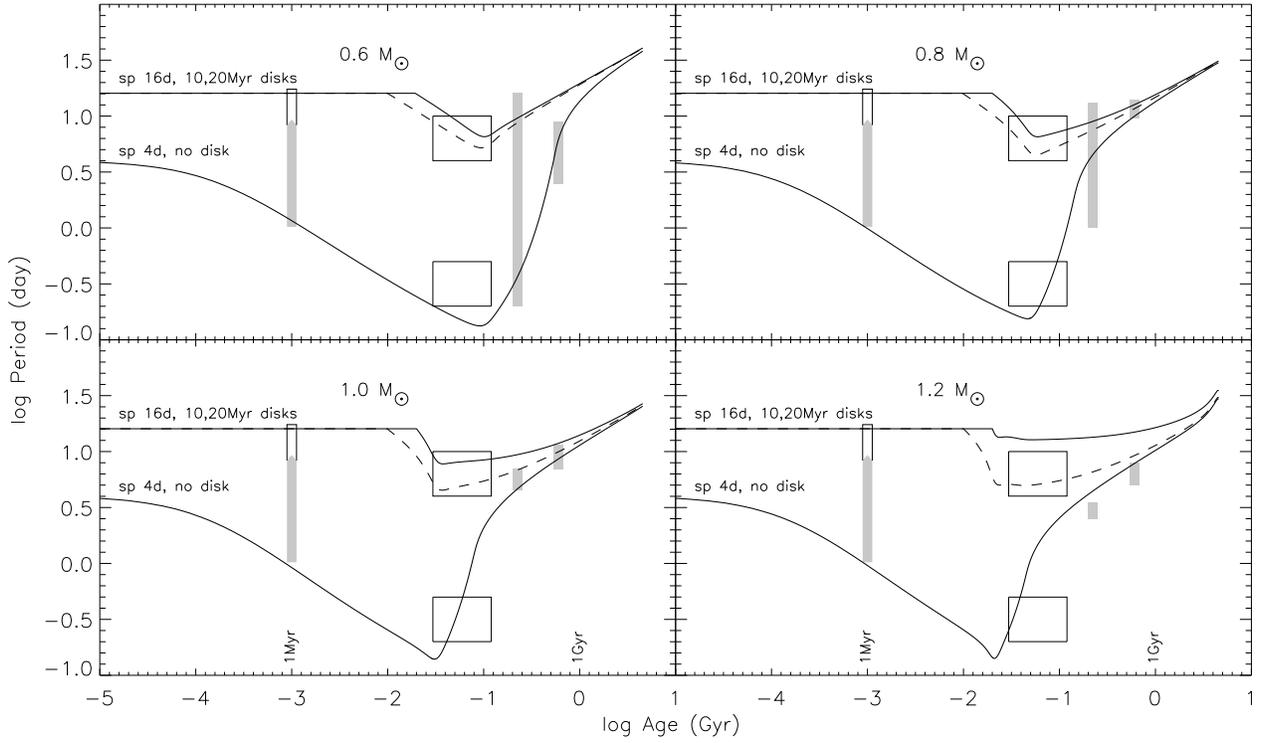}{6.5in}{0}{85}{85}{-290}{-150}
\caption[fig5.ps]{The range in stellar rotation period produced by solid-body models without disks and those disk-locked for 10 and 20 Myr. Note that with longer disk-locking timescales, the solid-body models can also explain the entire range of period observations in young clusters.}\label{fig5}
\end{figure}

\begin{figure}[6]
\plotfiddle{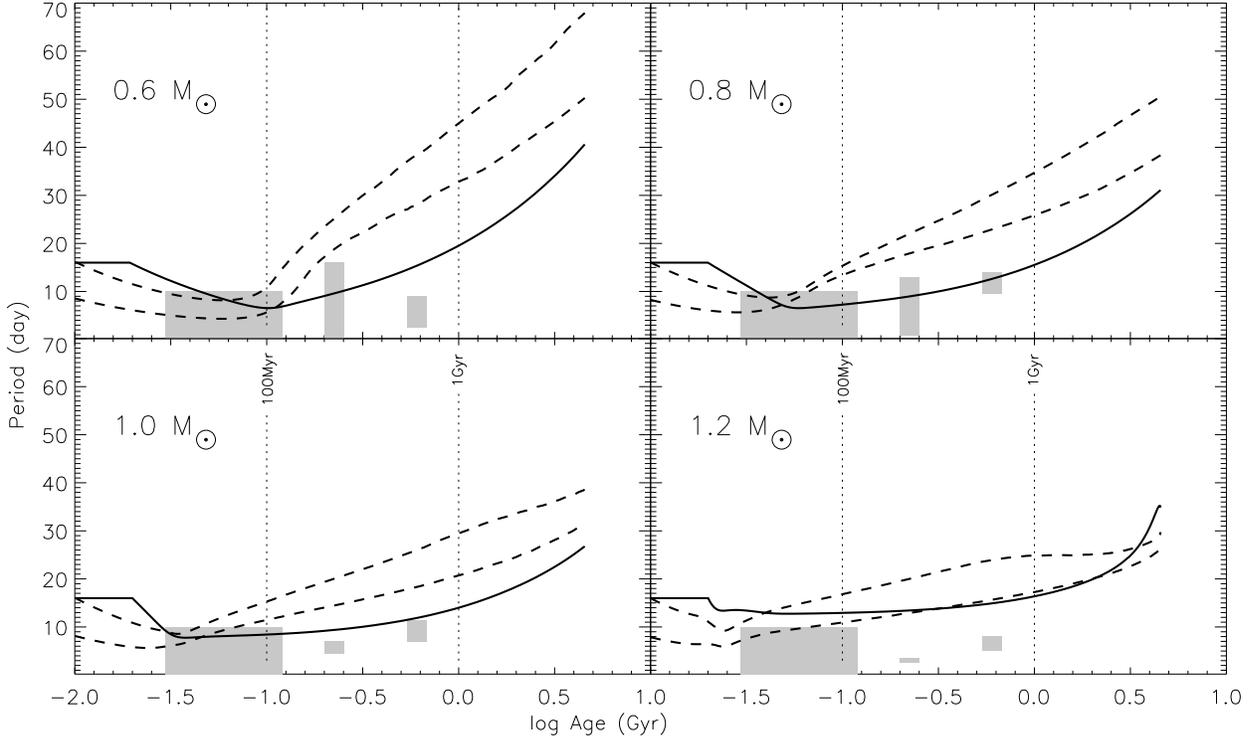}{6.5in}{0}{85}{85}{-290}{-150}
\caption[fig6.ps]{Comparison of the behavior of slow rotators in the solid-body and differentially rotating cases. The solid lines represent solid-body models with initial periods of 16 d and 20 Myr of disk-locking. The dashed lines represent differentially rotating models with the same initial period but 3 and 10 Myr of disk-locking. Note that the rotation period is plotted here on a linear scale. The shaded boxes in each panel represent the available observations in young clusters, NGC 3532 and the Hyades (Radick et al. 1987). If stars always rotate as solid bodies, we should not expect to find any objects above the solid lines in the figures.}\label{fig6}
\end{figure}

\end{document}